\documentstyle{mn}
\def\go{
\mathrel{\raise.3ex\hbox{$>$}\mkern-14mu\lower0.6ex\hbox{$\sim$}}
}
\def\lo{
\mathrel{\raise.3ex\hbox{$<$}\mkern-14mu\lower0.6ex\hbox{$\sim$}}
}
\def\simeq{
\mathrel{\raise.3ex\hbox{$\sim$}\mkern-14mu\lower0.4ex\hbox{$-$}}
}

\def\etal{{\it et al.\ }}

\def\etal{{\it et al.\ }}

\def\be{\begin{equation}}
\def\ee{\end{equation}}
\def\bea{\begin{eqnarray}}
\def\eea{\end{eqnarray}}

\def\etal{{\sl et al.\ }}

\def\hw2{{\hat W}^2}
\def\go{\mathrel{\raise.3ex\hbox{$>$}\mkern-14mu
             \lower0.6ex\hbox{$\sim$}}}
\def\lo{\mathrel{\raise.3ex\hbox{$<$}\mkern-14mu
             \lower0.6ex\hbox{$\sim$}}}
\def\ltorder{\mathrel{\raise.3ex\hbox{$<$}\mkern-14mu
             \lower0.6ex\hbox{$\sim$}}}
\def\gtorder{\mathrel{\raise.3ex\hbox{$>$}\mkern-14mu
             \lower0.6ex\hbox{$\sim$}}}

\def\eps2{{\epsilon^2}}

\input epsf.sty
\input psbox.tex

\begin{document}

\title[The highly variable X-ray spectrum of 1H~0419-577]
{The highly variable X-ray spectrum of the luminous Seyfert 1 galaxy
1H~0419-577}
\author[K.L. Page \etal]{K.L. Page, K.A. Pounds, J.N. Reeves and P.T.
O'Brien\\
X-Ray Astronomy Group, Department of Physics \& Astronomy,
Leicester, LE1 7RH, UK}

\date{Received ** *** 2001 / Accepted ** *** 2001}

\label{firstpage}

\maketitle

\begin{abstract}
An {\it XMM-Newton} observation of the luminous Seyfert 1 galaxy
1H~0419-577 is
presented. We find that the spectrum is well fitted by a power law
of canonical slope ($\Gamma$~$\sim 1.9$) and 3 blackbody components (to
model
the strong soft
excess).
The {\it XMM} data are compared and
contrasted with observations by {\it ROSAT} in 1992 and by {\it ASCA} and
{\it
BeppoSAX} in 1996. We find that the overall X-ray spectrum has changed
substantially over the period, and suggest that the changes are driven
by the soft X-ray component. When bright, as in our {\it XMM-Newton}
observation, it appears that the enhanced soft flux cools the
Comptonising corona, causing the 2--10 keV
power law to assume a `typical' slope, in contrast to the unusually
hard (`photon-starved') spectra observed by {\it ASCA} and {\it
BeppoSAX} four years earlier.

\end{abstract}


\section{Introduction}
\label{sec:intro}

1H~0419-577 (also known as LB~1727, 1ES~0425-573 and IRAS~F04250-5718)
is a radio-quiet
Seyfert galaxy, with a $60 \mu m$ flux of 0.18~Jy and an apparent
magnitude of 14.1. It is a moderate
redshift object ({\it z} $=0.104$) and relatively bright X-ray source
which has been observed over recent years by
{\it ASCA}, {\it ROSAT}
and {\it BeppoSAX}. 1H~0419-577 was also one of the
brightest Seyfert galaxies detected in the extreme ultra-violet
by the {\it ROSAT} Wide Field Camera (Pye \etal 1995) and
{\it EUVE} (Marshall, Fruscione \& Carone 1995).
Optical spectra taken over the same period in 1996 as the {\it ASCA}
and {\it BeppoSAX} X-ray observations (Guainazzi \etal 1998) show
1H~0419-577 to be a typical broad-line
Seyfert 1, in accordance with the classification by Brissenden (1989).

Over the 2--10~keV band, Seyfert galaxies can usually be modelled by a
power law,
with photon-index $\Gamma$ $\sim 1.8-2$.
Below about 1 keV a `soft
excess' is often reported, although
the limited bandwidth and resolution of
previous missions have made it difficult to distinguish a soft
{\it emission} component from the effects of absorption by ionised
matter. In the case of 1H~0419-577, Turner \etal (1999) did
conclude that
there is a soft
emission component on the basis of simultaneous
{\it ROSAT} HRI and {\it ASCA} observations. However,
those authors found the 2--10~keV power law to be unusually flat,
with $\Gamma$ $\sim 1.5-1.6$, and to extend down to 0.7~keV.
The {\it BeppoSAX} observation of 1H~0419$-$577 in 1996 September
(Guainazzi \etal 1998)
also found an unusually flat power law (over 3--10 keV) of
$\Gamma$ $\sim 1.55$, but provided no
independent
soft X-ray data, due to technical problems with the Low Energy
Concentrator Spectrometer (LECS).

The link between the hard X-ray power law and a soft X-ray emission
component is critical in the context of the accretion disc/corona model
for AGN, where the hard X-ray emission is explained by Comptonisation
of optical/EUV photons from the disc by energetic electrons in an
overlying corona (e.g., Haardt \& Maraschi 1991).
In this model the energy balance between the soft photon flux and
the corona then
determines the hardness of the spectrum in the 2--10~keV band.
The soft excess may be the
tail of the Big Blue
Bump, representing the thermal emission from an accretion disc
surrounding the central black
hole (Shields 1979; Czerny \& Elvis 1987; Ross \& Fabian 1993).

In our {\it XMM-Newton} observation, reported here, we find a
strong and broad soft emission component (consistent with an earlier
{\it ROSAT} PSPC observation in 1992; Guainazzi \etal 1998) and a 2--10~keV
power law
continuum slope typical of Seyfert 1 galaxies. We discuss our result
in terms of the stronger soft photon flux cooling the coronal electrons,
with a resulting steepening of the power law, and suggest it represents
direct observational support for the disc/corona model
for the hard X-ray emission from radio-quiet AGN.

In that general class of models, back-irradiation of the accretion disc
by the hard X-ray flux can result in
additional features being imprinted on the emerging X-ray spectrum.
The most obvious of these `reflection' features (Pounds \etal 1990; Nandra
\& Pounds 1994) is often an emission line at
6.4~keV arising from fluorescence in near-neutral Fe. This line has
emerged
as a powerful diagnostic of the inner regions in AGN since {\it ASCA}
observations found it to be broadened and red-shifted (Tanaka
\etal
1995; Nandra
\etal 1997a, b). Early observations from {\it XMM-Newton}
have shown a rather different situation to apply in several high-luminosity
Seyferts (similar to 1H~0419-577), with a weaker and higher energy (ionised)
broad Fe-K line, resolved from a narrow line at 6.4~keV (eg. Reeves
\etal 2001; Pounds \etal 2001). The latter component, interpreted
as scattering from neutral matter distant from the hard X-ray
source (e.g., in the molecular
torus),
is emerging as a common feature in AGN observations by {\it XMM-Newton}
and {\it Chandra}.

In previous observations of 1H~0419-577 Turner \etal (1999) found evidence
for an emission line in the
{\it ASCA} data at 6.39~keV,
with equivalent width (EW) of 700~$\pm$~400~eV (data from August
1996). A line is not
detected in the {\it SAX} data (Guainazzi \etal 1998), but
with a rather high upper limit of $\sim$~250~eV for the
equivalent width.

In Section~\ref{sec:xmmobs} the data from {\it XMM-Newton} are summarised,
followed by the data analysis in  Section~\ref{sec:specanal}.
Comparisons with the {\it ROSAT}, {\it ASCA} and {\it BeppoSAX}
observations are reviewed in Section~\ref{sec:otherobs}.
Note that all fit parameters are given for the rest frame of the
AGN, with values of $H_0$~=~50 km~s$^{-1}$~Mpc$^{-1}$ and
$q_0$~=~0 assumed throughout. Errors are quoted at the 90\% confidence
level ($\Delta \chi^{2}$~=2.7 for 1 interesting parameter).

\section{XMM-Newton Observations}
\label{sec:xmmobs}

The {\it XMM-Newton} observation of 1H~0419$-$577 took place on 4th
December 2000 and
lasted for just over 8~ks.
Because of a co-ordinate error, X-ray data were only obtained from the EPIC
PN camera (Str\"{u}der \etal 2001).

The PN data were reduced with the {\it
XMM} SAS (Science Analysis Software), using {\sc epchain} to produce
the event list. This was further filtered using {\sc xmmselect} within
the SAS. Both single and double pixel events (patterns 0-4 in {\sc
xmmselect}) were
selected, with the low energy cut-off being set to 200 eV. The
spectrum was extracted within a box-shaped region of size 20 by 10
pixels (corresponds to (87~$\times$~43.5)~arcsec).

The {\sc Xspec} v11.0 software package was used to analyse the
background subtracted spectrum, using the most recent response
matrices. (The spectrum was first binned, using the ftool command
{\it grppha}, to provide a minimum of 20 counts per bin.)

\section{Spectral Analysis}
\label{sec:specanal}

Since no significant changes occured in the X-ray flux over the
8000~second observation, the summed data were used in
modelling the spectrum.
The spectral analysis was begun in the conventional way,
comparing the data for 1H~0419-577 with a range of parametric models
until the best
reduced $\chi^{2}$ value was obtained.

First, a single absorbed power law ($N_h$ fixed to the Galactic
value of $2\times10^{20}$ cm$^{-2}$) was tried, but found to be
a very poor fit across the full 0.2--10~keV band
($\chi^2_{\nu} = 1409/712$), mainly due to a strong upward curvature
in the measured spectrum. Constraining the model to the
2--10~keV
band, however, provided a good fit for a photon-index $\Gamma$~$\sim~1.88$,
with $\chi^2_{\nu} = 325/349$.
A Gaussian component was then added to the model but yielded no significant
detection of either a narrow or a broad Fe
line. At a rest energy of 6.4~keV (appropriate to neutral iron
emission), the equivalent width of a narrow line ($\sigma$~=~10~eV) was
$<$~85~eV while for a broad neutral line ($\sigma$~=~300~eV) the upper limit
was
$<$~130~eV. For an
ionised line at 6.7~keV the corresponding broad line limit was
$<$~120~eV.
We note these upper limits are consistent with the weak Fe emission
lines detected in objects of similar luminosity (Reeves \etal 2001, Pounds
\etal 2001).

\begin{figure}
\psboxto(\hsize;0cm){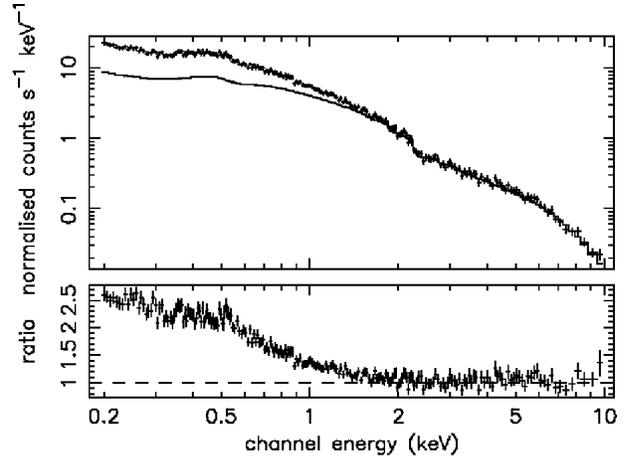}
\caption{Fit 1 for the PN spectrum, extrapolated down to 0.2~keV,
showing the soft excess of 1H~0419$-$577. The lower window shows the
ratio between the data and fitted model.}
\label{pn_softexcess}
\end{figure}

\begin{figure}
\psboxto(\hsize;0cm){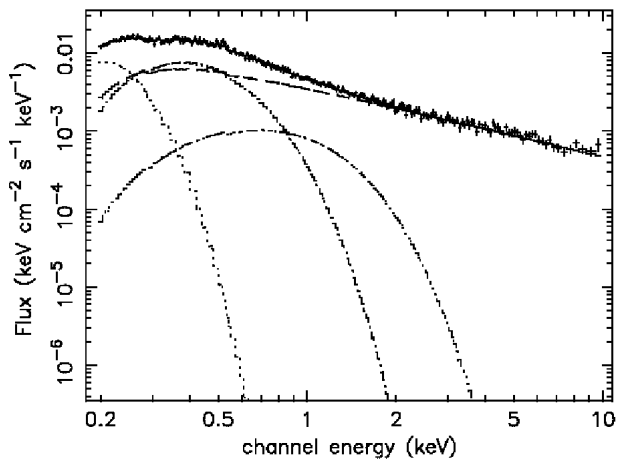}
\caption{Unfolded plot of the best fit to the {\it XMM} data (Fit 4 in
Table 2). The model
consists of a powerlaw ($\Gamma$~=~1.88) and 3 blackbody components
(kT~=~31~eV, 110~eV and 252~eV).}
\label{pn_unfolded}
\end{figure}

Extrapolating the above 2--10 keV power law fit down to 0.2~keV
revealed a
strong and broad (`hot') soft excess (Fig~\ref{pn_softexcess}).
The soft excess was then modelled by the addition of
blackbody emission, the breadth of the excess requiring 3 blackbody
components to match the observed spectrum (Fig~\ref{pn_unfolded}).
Details of this fit are given
as Fit 4 of Table 1.
The soft excess has no obvious superimposed absorption features, the limits for the
optical depths of the O~{\sc vii} and O~{\sc viii} edges being $\tau~<$~0.12
and $\tau~<$~0.06 respectively. A small excess of counts at
$\sim 0.55$~keV could be a calibration residual, although emission
from the O~{\sc vii} triplet has been previously identified, close to
this energy, in objects of similar luminosity (e.g., Mrk 359 (O'Brien \etal 2001),
Mrk 509 (Pounds \etal 2001)).
Without the benefit of RGS data, we
cannot investigate this further.

The measured flux of 1H~0419-577 at 2--10~keV was
1.12~$\times$~10$^{-11}$~erg~cm$^{-2}$~s$^{-1}$, corresponding to
a luminosity of
5.67~$\times$~10$^{44}$~erg~s$^{-1}$, with a flux of
2.88~$\times$~10$^{-11}$~erg~cm$^{-2}$~s$^{-1}$ over the full
0.2--10~keV band
(1.91~$\times$~10$^{45}$~erg~s$^{-1}$), including the strong 'soft
excess'.

\begin{table}
\begin{center}
\caption{Fits to the {\it XMM-Newton} data from 2000 December.}

\begin{tabular}{p{0.1truecm}p{1.0truecm}p{1.0truecm}p{1.5truecm}p{1.5truecm}p{1.0truecm}}
Fit & Range & Model & $\Gamma$ & kT & $\chi^{2}$/dof \\
& (keV) & & & (kev)\\
&\\
1 & 2--10 & PL & 1.88$\pm$0.03 & & 325/349 \\
&\\
2 & 0.2--10 & PL+BB & 2.05$\pm$0.01 & 0.108$\pm$0.002 &
839/710 \\
&\\
3 & 0.2--10 & PL+2BB & 1.99$\pm$0.02 & 0.033$\pm$0.002 &
 767/708 \\
& & & & 0.121$\pm$0.003 \\
&\\
4 & 0.2--10 & PL+3BB & 1.88$\pm$0.04 & 0.031$\pm$0.002 &
 734/706\\
 & & & & 0.110$\pm$0.005\\
 & & & & 0.252$\pm$0.030\\

\end{tabular}
\end{center}
\label{xmm}
\end{table}

\section{Comparison with other Observations}
\label{sec:otherobs}

The {\it XMM-Newton} spectrum of 1H~0419-577 is, on its own,
unremarkable, being very similar in overall continuum
shape to the recent {\it XMM-Newton} observations of the comparably
high-luminosity Seyfert 1 galaxies Mrk 205 and Mrk 509 (Reeves \etal 2001,
Pounds \etal 2001).
However, the {\it XMM-Newton} spectrum of 1H~0419-577 is dramatically different from
the several observations in 1996, which show a much flatter
2--10~keV continuum and weaker soft excess. To confirm this
spectral change, we have re-examined in a consistent way the
data from those previous missions, beginning with the {\it ROSAT} PSPC
observation in 1992, then those of {\it ASCA} in
July and August of 1996
(observation numbers 74056000 and
74056010) and in
September of that year by {\it BeppoSAX}. The {\it ROSAT} HRI also
monitored 1H~0419-577 simultaneously with {\it ASCA}, between 1996 June 30 and
September 1. The data from these
observations were retrieved from the LEDAS website
(http://ledas-www.star.le.ac.uk), the Tartarus Database
(http://tartarus.gsfc.nasa.gov) and the {\it SAX} homepage
(http://brunello.sdc.asi.it) respectively. Only the data from the SIS
instruments on {\it ASCA} were used.

\begin{figure}
\psboxto(\hsize;0cm){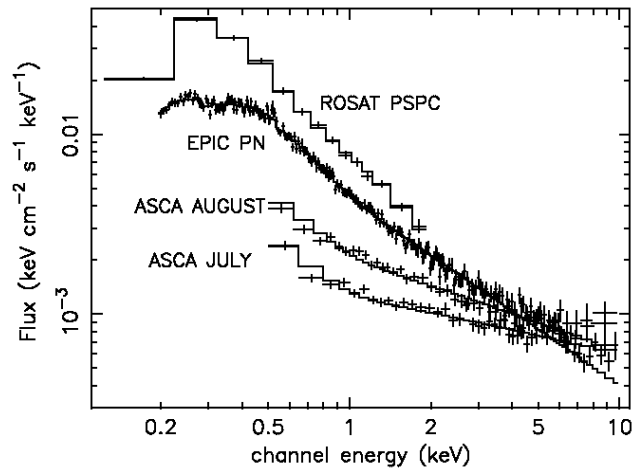}
\caption{Plot showing the difference between the unfolded spectra for
the {\it XMM}, {\it ASCA} and {\it ROSAT} observations. The model used
was a powerlaw and 2 BBs. It can clearly be seen that the {\it
ROSAT} and {\it XMM} datasets have steeper powerlaw slopes and
stronger soft X-ray emission than
the {\it ASCA} observations.}
\label{asca+xmm}
\end{figure}

We began by fitting a power law plus Galactic absorption to
the {\it ASCA} and {\it SAX} datasets over the 2--10~keV band.
In each case the power-law slope was significantly flatter compared to {\it XMM}. Details of
these power-law fits are given in
Table~2. It should be noted that we have fitted
over the entire 2--10~keV band, whereas Turner \etal (1999) ignored
5.0-7.5~keV within this range, where iron emission would occur if present.
If we do ignore this region, then the {\it ASCA} slopes steepen by approximately
0.1, agreeing more closely with those derived by Turner \etal
Interestingly, although the slopes are very different, the 2--10~keV
flux remains essentially constant
between the {\it XMM}, {\it ASCA} and {\it BeppoSAX}
observations. This is a point
also illustrated in
Figure~\ref{asca+xmm}.

\begin{table*}
\begin{center}
\caption{Fits to the {\it ASCA}, {\it SAX} and {\it XMM} data.  $^f$ BB components frozen to PN values, with normalisations tied to
the same ratio}
\begin{tabular}{p{2.5truecm}p{0.5truecm}p{1.5truecm}p{2.5truecm}p{2.5truecm}p{2.0truecm}p{2.0truecm}p{2.0truecm}p{2.0truecm}p{2.0truecm}}

\large\bf{2--10~keV}\\
&\\
Mission & Fit & Model & $\Gamma$ & kT~(keV) &
$\chi^{2}$/dof & Flux~(erg~cm$^{-2}$~s$^{-1}$) \\
&\\
ASCA (1996 July) & 1 & PL & 1.35 $\pm$ 0.05 && 162/134 & 9.32~$\times$~10$^{-12}$\\

&\\
ASCA (1996 Aug.) & 1 & PL & 1.42 $\pm$ 0.06 & & 91/140 & 1.13~$\times$~10$^{-11}$\\

&\\
SAX (1996 Sept.)& 1 & PL & 1.63 $\pm$ 0.04 & & 140/111 & 1.15~$\times$~10$^{-11}$\\
&\\

\large\bf{0.5--10~keV}\\
&\\
XMM & 5 & PL+2BB & 1.87 $\pm$ 0.03 & 0.093 $\pm$ 0.006 &
661/650 & 2.20~$\times$~10$^{-11}$ \\
& & &  & 0.225 $\pm$ 0.020\\
&\\
ASCA (1996 July) & 5 & PL+2BB & 1.32 $\pm$ 0.03 & 0.093$^f$ &
222/185 & 1.27~$\times$~10$^{-11}$ \\
& & & & 0.225$^f$\\
&\\
ASCA (1996 Aug.)& 5 & PL+2BB & 1.53 $\pm$ 0.03 & 0.093$^f$ &
143/191 & 1.61~$\times$~10$^{-11}$ \\
&  & & & 0.225$^f$\\

\end{tabular}
\end{center}
\label{fitcomparisons}
\end{table*}

The PN spectrum was then re-fitted over the 0.5--10~keV {\it ASCA} band, so that
the {\it ASCA} and {\it XMM-Newton} data could be compared
directly. Over this band, the PN data are well fitted with a power law
and only 2
blackbody components. Each {\it ASCA}
observation was then modelled in this way, with the BB temperatures and
relative normalisations frozen to the values determined from the {\it
XMM} data, since the soft excess is not well constrained for the {\it
ASCA} datasets. The results of this fit (Fit 5) are given in
Table~2. The 0.5--2~keV flux is a factor of $\sim~3$ higher in the
{\it XMM-Newton} observation and the power law is much steeper ($\Delta
\Gamma~\sim~0.4$). The power law slope and soft excess change in the
same sense, but to a lesser degree, between the two {\it ASCA} observations, with the soft flux
increasing by 42\% from July to August 1996 according to Fit 5, in
excellent agreement with the $\sim 40$\% increase in count-rate
recorded by the {\it ROSAT} HRI data (Turner \etal 1999).

Since only the data from the MECS instrument onboard
{\it BeppoSAX} were
available, BB components could not be constrained for that observation.
The power law slope is 1.63~$\pm$~0.04, implying that the slope was
continuing to steepen, as hinted at between the {\it ASCA} observations.

To compare the flux between {\it ROSAT} and {\it XMM}, a simple power law was fitted to the
{\it ROSAT} and PN datasets, over
the range 0.2--2.0~keV, with $N_h$ fixed to the Galactic value.
 For {\it ROSAT}, $\Gamma$ $\sim 2.94$, while
the PN slope was slightly less steep, at $\Gamma$ $\sim 2.46$.
This corresponds to a
flux over the 0.2--2~keV band of 3.42~$\times$~10$^{-11}$~erg~cm$^{-2}$~s$^{-1}$  for
{\it ROSAT}, compared to
1.76~$\times$~10$^{-11}$~erg~cm$^{-2}$~s$^{-1}$ as measured by {\it
XMM}.
We stress that a  single power law is not a good fit to either the
{\it ROSAT} or {\it XMM} data, both of which show evidence for a more
complex soft excess.
It has been noted by a number of authors (e.g., Turner 1993; Iwasawa,
Brandt \& Fabian 1998) that the {\it ROSAT} PSPC spectral indices are
up to 0.4 steeper than those found from other missions. Allowing for
these uncertainties, we consider that the {\it ROSAT} PSPC observation of 1992 is consistent
with the soft
excess at that time being as strong as in the 2000 December {\it
XMM-Newton} observation.

The next step was to compare the {\it ASCA}, {\it SAX} and {\it XMM}
data to search for neutral iron emission. We found no evidence for a  narrow
line
($\sigma$~=~10~eV) in any of the observations. The July 1996 {\it ASCA}
data gave an equivalent width of $<$~81~eV, very similar to our {\it XMM} upper limit.
The August
{\it ASCA} and September {\it BeppoSAX} data yielded
upper limits of $\sim$140~eV. Broad line fits (width 300~eV)
also gave only upper limits of order $<$~350~eV for the July
{\it ASCA} and September {\it BeppoSAX} data, while for the  August {\it
ASCA} data the fit did show a decrease in $\chi^2$ of
7 for 1 additional degree of freedom (a confidence level of
greater than 99\%). This line, if real, has an EW of 335~$\pm$~210~eV.
Summarising, the August 1996 {\it ASCA} data suggest that a broad iron line
may exist, although none of the other observations detected such emission.

It is known that the low energy response of the {\it ASCA} SIS has
been degrading and that this leads to an underestimate of the soft
X-ray flux.
In order to estimate the decrease in the low energy
efficiency of {\it ASCA} in the 1H~0419-577 observation, we used the
formula given in the {\it ASCA} GOF Calibration Memo (Yaqoob \etal 2000),
which quotes the effective increase in N$_h$ as a
function of time. The net effect of this in our fits is a change of
$\Gamma$$\sim~0.05$ in the spectral slope. This does not,
therefore, alter our conclusions about the change in state of
1H~0419-577
between the {\it ASCA} and {\it XMM} observations.

\section{Discussion}
\label{sec:disc}

Our {\it XMM-Newton} observation of 1H~0419-577 found the source in a
bright/soft state. The overall 0.2--10~keV spectrum is
very similar to that of 2 other luminous Seyfert galaxies recently observed with
{\it XMM-Newton}, Mrk 205 (Reeves \etal 2001) and Mrk 509 (Pounds \etal
2001), exhibiting a typical 2--10~keV power law slope and strong soft
excess.
However,
the new observation of 1H~0419-577 contrasts markedly with both
{\it ASCA} and {\it BeppoSAX} spectra obtained in 1996, where the
2--10~keV continuum was significantly flatter, and the soft excess
much weaker. We suggest the strong soft excess
in our recent observation is the key to the overall spectral change,
with increased cooling of the coronal
electrons yielding a steeper/softer hard power law component. Interestingly, it seems possible that this effect was
already apparent between the 2 {\it ASCA} observations a month apart
in 1996, with an increase in the {\it ROSAT} HRI flux of at least
40\%,
and the best fit 2--10~keV power law slopes increasing from
1.35~$\pm$~0.05 to
1.42~$\pm$~0.06 (see also Turner \etal 1999).
If the 2000 December spectral state is considered `normal' for a BLS1
then,
in the framework of the disc/corona model, we can
attribute the unusually flat/hard 2--10~keV continuum in the 1996
observations to
a lack of soft photons (`photon starved' Comptonisation).

In order to quantify such a change we have tried a Comptonisation fit to
the data ({\it comptt} in {\sc Xspec}), using a model (as in O'Brien
\etal 2001) in which soft photons from the accretion disc
are up-scattered by thermal electrons characterised by two
temperatures, to yield
the observed broad soft excess and the harder power law respectively.
For the {\it XMM-Newton} data we assumed an input photon distribution
of kT~=~20~eV (appropriate to thermal radiation from the inner disc
of a 10$^8$ $M_{\odot}$ black hole).

\begin{figure}
\psboxto(\hsize;0cm){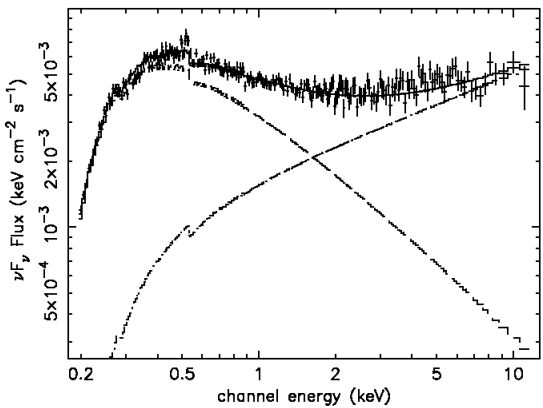}
\caption{Best fit Comptonisation model for the EPIC PN dataset. Input photons at kT~=~20~eV are Comptonised by electron distributions of kT~=~2.5~keV and 95~keV.}
\label{comptt}
\end{figure}

For the {\it XMM-Newton} data we obtained an acceptable fit
($\chi^2_{\nu}~=~1.11$) across the
full 0.2--10 keV spectrum with
a `cool' Comptonising component of kT~=~2.5~$\pm$~0.1 keV and optical depth
of $\tau$~=~3.9~$\pm$~0.1, together with a `hot' component of kT~=~95~$\pm$~15 keV and
$\tau$~=~1.0~$\pm$~0.1
(Figure~\ref{comptt}).
The same double-Comptonisation model was then compared with the {\it
ASCA} data from the 1996 July observation. Again an acceptable fit was
obtained, with
Comptonising components of kT~=~2.5~$\pm$~0.1 keV with an optical depth
of $\tau$~=~3.5~$\pm$~0.1 (and much smaller normalisation)
and kT~=~168~$^{+100}_{-30}$ keV
($\tau$~=~1.0~$\pm$~0.1). While the individual parameters should not be taken too
literally, since there is a close coupling of temperature and optical
depth, the fits
support the view that the steepening of the power law
between the 1996 and 2000 observations of 1H~0419-577 follows a significant
cooling of the hot Comptonising electrons. This is consistent with the much
stronger soft X-ray flux detected in the {\it XMM-Newton} observation.

Finally we note that the relatively short exposure of the {\it XMM}
observation means that the limits on the Fe K lines are not very
severe. However, it might be expected that any broad line would
be weakened if the disc had an optically thick, ionised
skin, as our Comptonisation model suggests.

\section{Conclusions}

A new measurement of the X-ray spectrum of the luminous Seyfert 1 galaxy
1H 0419-577 with {\it XMM-Newton} has shown it to have a strong soft excess
and `normal' power law slope, very similar to the comparably luminous
Seyferts Mrk 205 and 509. No iron line was found, but the upper limits
are consistent with the Fe-K emission found in {\it XMM-Newton}
observations of objects of similar luminosity.
A comparison with archival data shows
the 2--10~keV continuum to have been much flatter in 1996, together with a far
weaker soft excess. We interpret our overall {\it XMM-Newton}
observation in terms of two-temperature Comptonisation of thermal
photons from the
inner accretion disc, and explain the hard continuum seen in 1996 as
being due
to `photon starving' of the hotter electron component.
We note the spectral changes we have observed in the Seyfert galaxy
1H 0419-577 are similar to, but smaller than, those seen in the
changes in `state' for several galactic black hole candidate sources
(e.g., Vilhu \etal 2001).

\section{ACKNOWLEDGMENTS}
The work in this paper is mainly based on observations with {\it
XMM-Newton}, an ESA
science mission, with instruments and contributions directly funded by
ESA and NASA. The authors would like to thank the EPIC Consortium, led
by Dr M.J.L. Turner, for all their work during the calibration phase,
and the SOC and SSC teams for making the observation and analysis
possible.
This research has also made use of the NASA/IPAC Extragalactic
Database (NED), which is operated by the Jet Propulsion Laboratory,
California Institute of Technology, under contract with the National
Aeronautics and Space Administation.
Support from a PPARC studentship and the Leverhulme Trust is
acknowledged by KLP and JNR respectively.

\end{document}